\documentclass[aps,prb,twocolumn,showpacs]{revtex4}
\usepackage{epsfig}
\usepackage{prettyref}
\begin {document}
\title {Signatures of Spin Glass Freezing in NiO Nanoparticles}
\author{S. D. Tiwari}
\affiliation{Department of Physics, Indian Institute of Technology Kanpur 208016, India}
\author{K. P. Rajeev}
\affiliation{Department of Physics, Indian Institute of Technology Kanpur 208016, India}
\begin{abstract}
We present a detailed study of the magnetic properties of sol-gel prepared nickel oxide nanoparticles of different sizes. We report various measurements such as frequency, field and temperature dependence of \emph{ac} susceptibility, temperature and field dependence of \emph{dc} magnetization and time decay of thermoremanent magnetization. Our results and analysis show that the system behaves as a spin glass. 
\end{abstract}

\pacs{61.46.+w, 75.20.-g, 75.50.Ee, 75.50.Tt, 75.50.Lk, 75.75.+a}


\maketitle

\section{Introduction}
For the past few decades magnetic nanoparticles have been attracting the attention of scientists from diverse disciplines from the standpoints of both fundamental understanding and useful applications \cite{Chantrell}. Magnetic nanoparticles have been under scrutiny since the days of N\'{e}el \cite{Neel} and Brown \cite{Brown} who developed the theory of magnetization relaxation for noninteracting single domain particles. Effect of inter-particle interaction on magnetic properties of several nanoparticle systems are available in the literature \cite{Andersson}. In 1961 N\'{e}el suggested that small particles of an antiferromagnetic material should exhibit superparamagnetism and weak ferromagnetism \cite{Low Temp. Phys.}. Antiferromagnetic nanoparticles are known to show more interesting behavior compared to ferro and ferrimagnetic nanoparticles one of which is that the magnetic moment of tiny antiferromagnetic particles increases with increasing temperature \cite{Seehra 2000, Harris, Kilcoyne, Makhlouf}. This is quite unlike what is seen in ferro and ferrimagnetic particles. Real magnetic nanoparticles have disordered arrangement, distribution in size and random orientation of magnetization, making their behavior very complex and challenging to understand.

Bulk NiO has a rhombohedral structure and is antiferromagnetic below 523~K whereas it has a cubic structure and is paramagnetic above that temperature \cite{Smart}. A bulk antiferromagnet has zero net magnetic moment in zero applied field. If the surface to volume ratio, which varies as the reciprocal of the particle size, for an antiferromagnetic particle becomes sufficiently large then the particle can have a nonzero net magnetic moment because of uncompensated spins at the surface. According to N\'{e}el the moment due to uncompensated spins would be parallel to the axis of antiferromagnetic alignment. If we consider a collection of such particles of varying sizes and shapes we will see that there will be a distribution of magnetic moments of different sizes, oriented randomly and interacting with each other magnetically.  Thus the magnetic properties of a collection of such particles would be very different from that of the corresponding bulk material. Richardson and Milligan \cite{Richardson and Milligan} were the first to report magnetic susceptibility measurements as a function of temperature for NiO nanoparticles of different sizes. They measured the magnetic susceptibility at a field of 3500~G and found a peak in the susceptibility much below the N\'{e}el temperature and the peak temperature was found to decrease with decreasing particle size. They attributed the observed behavior to the modified magnetic environment of the Ni$^{2+}$ ions at the surface of the particles and the enhanced surface to volume ratio. 

In this paper we present a detailed study on NiO nanoparticles based mainly on magnetic measurements. In the course of this work we will come across phenomena  which indicate that a collection of NiO nanoparticles behaves as a spin glass or a superparamagnet. In each of those cases we shall try to critically examine the data in the light of current understanding of the physics involved. 
 
\section{ Experimental Details}
NiO nanoparticles are prepared by a sol-gel method by reacting in aqueous solution, at room temperature, nickel nitrate and sodium hydroxide at pH = 12 as described elsewhere \cite{Richardson and Milligan, Makhlouf 1997, Richardson 1991}. In this work we used nickel (II) nitrate hexahydrate   (99.999\%), sodium hydroxide pellets (99.99\%), both from Aldrich, and triple distilled water to make nickel hydroxide. The samples of nickel oxide nanoparticles are prepared by heating the nickel hydroxide at a few selected temperatures for 3 hours in flowing helium gas (99.995\%). All the magnetic measurements are done with a SQUID magnetometer (Quantum Design, MPMS XL).
\section{Experimental Results and Discussion}
\subsection{Crystallite sizes}
The average crystallite size is calculated by X-ray diffraction line broadening using the Scherrer formula \cite{Scherrer}
\begin{eqnarray}
t = \frac{0.9 \lambda}{\cos \theta_{B} \sqrt{B_{M}^{2}-B_{S}^{2}}}
\end{eqnarray}
where $\lambda$ is the wavelength of the X-ray (1.542 \AA), 2$\theta_{B}$ is the Bragg angle, $B_{M}$ is the full width at half maximum (FWHM) of a peak in radians and $B_{S}$ is the FWHM of the same peak of a standard sample. We used specpure grade NiO powder from Johnson Matthey \& Co. Ltd. (UK) as the standard. Peaks (111) , (200) and (220) are used to calculate the average crystallite size. The use of \( \sqrt{B_{M}^{2}-B_{S}^{2}} \) instead of $B_{M}$ in the Scherrer formula takes care of instrumental broadening. The crystallite sizes of NiO samples prepared by heating Ni(OH)$_{2}$ at 250, 300, 350 and 700 $^{\circ}$C turn out to be 5.1, 6.2, 8.5 and $> 100$~nm respectively and these numbers will be  referred to as the average crystallite size in this paper. Transmission electron micrograph of NiO sample prepared by heating Ni(OH)$_{2}$ at 250 $^{\circ}$C gives mean particle size to be 5.6~nm with a standard deviation of 1.3~nm. We note that the mean particle size determined by transmission electron micrograph is very close to the average crystallite size determined by X-ray diffraction using the Scherrer formula (5.1~nm) which implies that on an average each NiO nanoparticle is a crystallite (tiny single crystal). More details of sample characterization like X-ray diffraction pattern, transmission electron micrograph, selected area electron diffraction pattern and particle size distribution are available elsewhere \cite{arxiv} .

\subsection{\emph{ac} Susceptibility} 
\label{ac-chi}

\subsubsection{Temperature and Frequency Dependence} 
\label{subsec:T&Fdepend-ac-chi}

\begin{figure}[ht]
\begin{center}
\includegraphics[angle=0,width=1.\columnwidth]{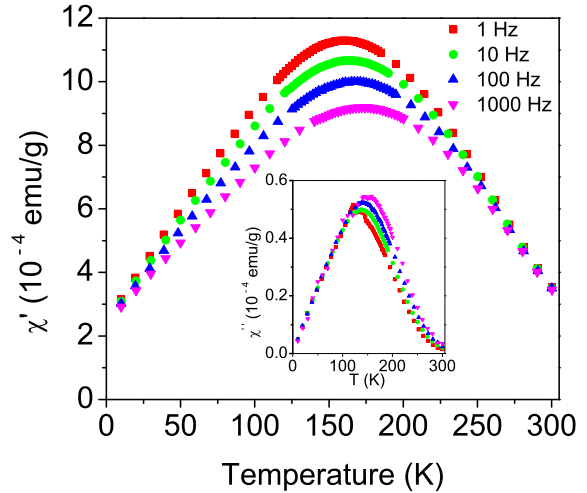}
\caption{(color online) Temperature and frequency dependence of real part of \emph{ac} susceptibility of 5.1~nm NiO nanoparticles. The inset shows the temperature and frequency dependence of the imaginary part of the \emph{ac} susceptibility. The probing field has an amplitude of 1~G.}
\label{fig:ac-suscept-frequency}
\end{center}
\end{figure}

The temperature dependence of \emph{ac} susceptibility is measured at several frequencies ranging from 1 Hz to 1 kHz. The data are taken as described below. The sample is first cooled from room temperature to 10~K in a zero magnetic field. Then a probing \emph{ac} magnetic field of 1.0~G amplitude is applied to measure the susceptibility as the temperature is slowly raised in short steps to 300~K. Figure \ref{fig:ac-suscept-frequency} shows the real part, $\chi'$, of the $ac$ susceptibility of the 5.1~nm sample; in the inset we show the imaginary part, $\chi''$. We note that all the curves have a peak at some temperature; as the frequency is raised the value of $\chi'$ decreases and the temperature of the peak increases. The behavior observed here is characteristic of both superparamagnets and spin glasses. The cusp in the $ac$ susceptibility seen in canonical spin glasses is not observed here. The inset shows that below the peak temperature, $\chi''$ does not depend on frequency. Such frequency independent $\chi''$ has been observed in other nanoparticle systems as well\cite{Jonsson}.

A quantitative measure of the peak temperature shift with frequency is the relative shift in peak temperature, ${\Delta T_{f}}/{T_{f}}$, per decade of frequency. For the 5.1~nm sample this quantity turns out to be 0.018. For many canonical spin glasses it lies between 0.0045 and 0.06 whereas for a known superparamagnet a-(Ho$_2$O$_3$)(B$_2$O$_3$) a value of 0.28 has been reported \cite{Mydosh}. Ferritin, an antiferromagnetic nanoparticle, which has been found to be  superparamagnetic, has a value of $\approx 0.13$ for this quantity\cite{Kilcoyne}.  We note that our value of 0.018 falls in the spin glass range.

Small particles of antiferromagnetic materials are expected to be superparamagnetic just as small particles of ferro or ferrimagnetic materials. That is, each particle would behave as a single giant moment and their magnetic susceptibility as a function of temperature would be Curie-like at sufficiently high temperature. The giant moment arises because of uncompensated spins at the surface of the particle.  An antiferromagnetic particle with uni-axial anisotropy has two low energy states, separated by an energy barrier $E_a$, corresponding to the parallel or anti parallel alignment of the magnetization of the particle with the axis of antiferromagnetic alignment\cite{Bean}.  The susceptibility has a maximum at a certain temperature called the blocking temperature, below which the probability of thermally assisted transitions between the two low energy states decreases progressively towards zero. The blocking temperature, $T_B$, should increase with increasing particle size because the energy barrier separating the low energy states is proportional to the volume of the particle. In fact
\begin{equation}
\label{Super-para-blocking}
T_B \propto V(H_K-H)^2
\end{equation}
where $V$ is the volume of a single particle, $H_K$ is a constant and $H$ is the field of measurement\cite{Bitoh}.

The dynamics of superparamagnets is described by Arrhenius law 
\begin{equation}
\nu = \nu_{0} \exp(-E_{a}/k_{B}T)
\end{equation}
where $E_{a}$ is the energy barrier for flipping the magnetic moment of the particle and $\nu_{0}$, an attempt frequency, usually has a value in the range $10^{8}$ to $10^{12}$ Hz and is related to the intra-potential-well dynamics\cite{Labarta, Morup 1995}. If we fit the data shown in Figure \ref{fig:ac-suscept-frequency} to the Arrhenius equation, where $T$ corresponds to the peak in the susceptibility curve taken at frequency $\nu$, we get $E_{a} \approx 14500$~K and $\nu_{0} \approx 10^{39}$ Hz; these numbers are too large and rather unphysical. Unreasonable numbers like these are usually seen in spin glass systems\cite{Mydosh} while superparamagnets tend to give more reasonable numbers such as $E_{a} \approx 300$~K and $\nu_{0} \approx 10^{11}$ Hz for ferritin\cite{Kilcoyne}. Here we have another indication that the NiO particles are not behaving like a superparamagnetic system in the region of temperature where they have a susceptibility peak.  This fact and the previously noted peak temperature shift with frequency suggest that the maximum in the \emph{ac} susceptibility curve might have its origin in spin glass like freezing rather than superparamagnetic blocking.

\subsubsection{Particle Size Dependence} 
\label{subsec:ac-suscept-particle-size}

\begin{figure}[ht]
\begin{center}
\includegraphics[angle=0,width=1.\columnwidth]{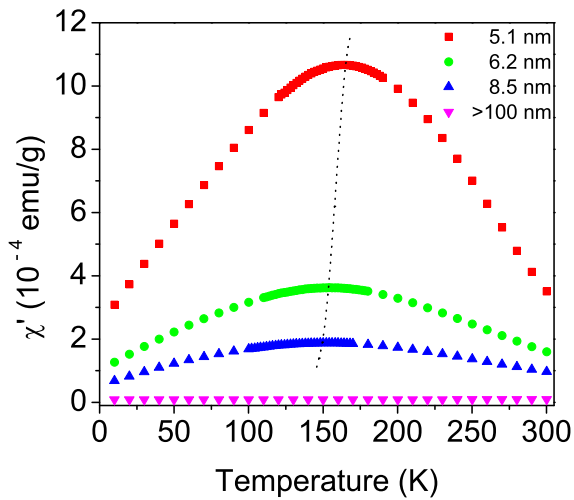}
\caption{(color online) Temperature variation of the real part of \emph{ac} susceptibility of NiO nanoparticles in an $ac$ field of 1~G amplitude  and 10~Hz frequency. The dotted line drawn to pass through the peaks shows that the peak temperature increases with decreasing particle size.}
\label{fig:ac-suscept-particle-size}
\end{center}
\end{figure}

In Figure \ref{fig:ac-suscept-particle-size} we compare $\chi'$  for 5.1, 6.2, 8.5 and $>$~100~nm  samples at 10 Hz. It is clear that the temperature of the peak decreases with increasing particle size. This behavior is quite contrary to what is expected in superparamagnetic systems as described by Equation \ref{Super-para-blocking} where $T_B \propto$ volume of particle. We see that just as the frequency dependence of the peak temperature was unlike that of a superparamagnet, its size or volume dependence is also unlike that of a  superparamagnet. In this paper we make an attempt to understand this non-intuitive behavior of NiO nanoparticles.

We also note from Figure \ref{fig:ac-suscept-particle-size} that the magnitude of the susceptibility increases with decreasing particle size. This is as one would expect since the relative number of surface spins and hence the number of uncompensated surface spins would increase with decreasing particle size. Our data is in qualitative agreement with the claim that in NiO nanoparticles susceptibility at room temperature varies as $1/d$ where $d$ is the particle diameter\cite{Richardson 1991}. Owing to the too few number of data points in our data set we are not able to make any quantitative claim.

\subsubsection{Magnetic Field Dependence}
\label{subsec:Field-depend-ac-chi}

\begin{figure}[t]
\begin{center}
\includegraphics[angle=0,width=1.\columnwidth]{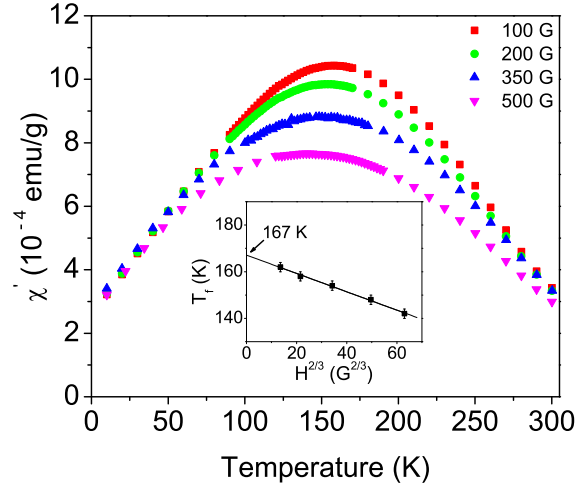}
\caption{(color online) Temperature variation of the real part of \emph{ac} susceptibility of 5.1~nm NiO nanoparticles at different \emph{dc} fields. $T_{f}$ are plotted against $H^{2/3}$ in the inset. The solid line shows a linear fit to the data with coefficient of determination $R^2=0.9966$.}
\label{fig:ac-suscept-field}
\end{center}
\end{figure}

We measured the $ac$ susceptibility at various bias fields $H$ with an $ac$ field of 1~G amplitude and 10 Hz frequency. This is shown in Figure \ref{fig:ac-suscept-field}. As $H$ increases, the peak temperature and peak value decreases. We note that the peak temperature, $T_{f}$, decreases as, $\delta T_{f} \propto H^{2/3}$, as shown in the inset of Figure \ref{fig:ac-suscept-field}.  This dependence corresponds to the so called de Almeida-Thouless (AT) line \cite{AT} given by: 
\begin{eqnarray}
H \propto (1-T/T_{f})^{3/2}
\end{eqnarray}
The extrapolation of the AT line back to $H = 0$ gives the spin glass transition temperature $T_{f}$ which, in this case, turns out to be about 167~K. Compliance of the data with the AT line is considered to be a strong evidence for the existence of a spin glass phase and it has been observed in different kinds of spin glasses \cite{Chamberlin 1982}. Recently, it has been used as an evidence for spin glass phase in $\gamma-Fe_{2}O_{3}$ nanoparticles \cite{Martinez} as well as in thin films \cite{Dhar}.  

\subsection{\emph{dc} Magnetization}
\subsubsection{Temperature Dependence}

\begin{figure}[hbt]
\begin{center}
\includegraphics[angle=0,width=1.\columnwidth]{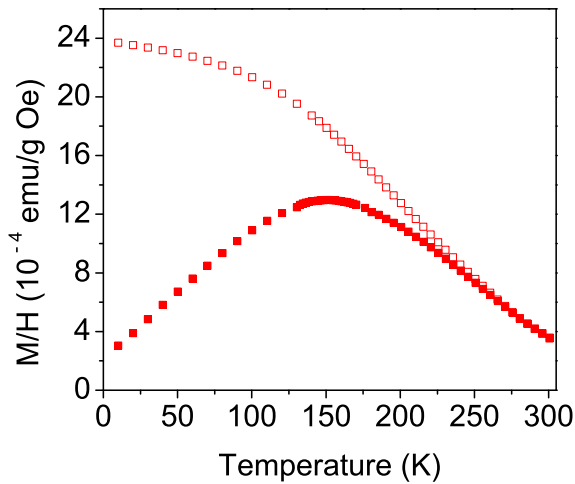}
\caption{(color online) Magnetization, ZFC (solid symbol)\ and FC (open symbol)\ for 5.1~nm NiO particles as a function of temperature in 100~G field.}
\label{fig:dc-mag-temp}
\end{center}
\end{figure}

Figure \ref{fig:dc-mag-temp} shows the temperature dependence of zero field cooled (ZFC) and field cooled (FC) magnetization in a 100~G $dc$ field for the 5.1~nm particles. Generally, the ZFC susceptibility shows a peak for both superparamagnets and spin glasses. In contrast, it is usually seen that the temperature dependence of the FC susceptibility becomes saturated below the peak temperature (freezing temperature $T_{f}$) for  spin glasses and continues to increase below that temperature (blocking temperature $T_{B}$) for superparamagnets \cite{Bitoh}. Nevertheless, we note that, glassy behavior in magnetic nanoparticles has also been claimed in which the FC susceptibility continues to increase with decreasing temperature \cite{Luo}. In our case, below the peak temperature, the susceptibility continues to increase but with a tendency towards saturation. This behavior is rather ambiguous.

\subsubsection{Particle Size Dependence}
\label{subsec:dc-mag-particle-size}

\begin{figure}[hbt]                                                             
\begin{center}                                                                  
\includegraphics[angle=0,width=1.\columnwidth]{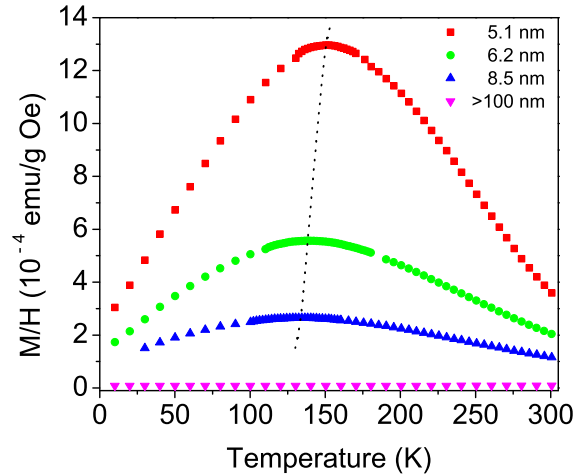}
\caption{(color online) Magnetization (ZFC) as a function of temperature in 100~G applied field for NiO nanoparticles of various sizes. The dotted line, drawn to pass through the peaks, shows that the peak temperature decreases with increasing particle size.}
\label{fig:dc-mag-particle-size}
\end{center}                                                                    
\end{figure}

Figure \ref{fig:dc-mag-particle-size} shows our data on the temperature dependence of ZFC magnetization as a function of particle size in an applied field of 100~G.  Clearly the peak temperature decreases with increasing particle size, as has been seen in \emph{ac} susceptibility measurements in Section \ref{subsec:ac-suscept-particle-size} earlier. Again, we notice the non-superparamagnetic behavior. A similar observation has been reported recently for MnO nanoparticles where the temperature of the peak in $dc$ susceptibility in a zero field cooled measurement decreases with increasing particle size \cite{Seo}.

The behavior we see here at 100~G field does not agree with the observations reported by Richardson and Milligan\cite{Richardson and Milligan} on NiO where they found that at 3500~G the peak temperature in \emph{dc} susceptibility increases with increasing particle size, as already mentioned in the Introduction. We did $dc$ susceptibility measurement at 3500~G and found that we are able to reproduce\cite{arxiv} the data of Richardson and Milligan. We shall see later in Section \ref{Further Discussion} that these two conflicting observations are actually consistent with each other.

\subsubsection{Field Dependence}

\begin{figure}[hbt]
\begin{center}
\includegraphics[angle=0,width=1.\columnwidth]{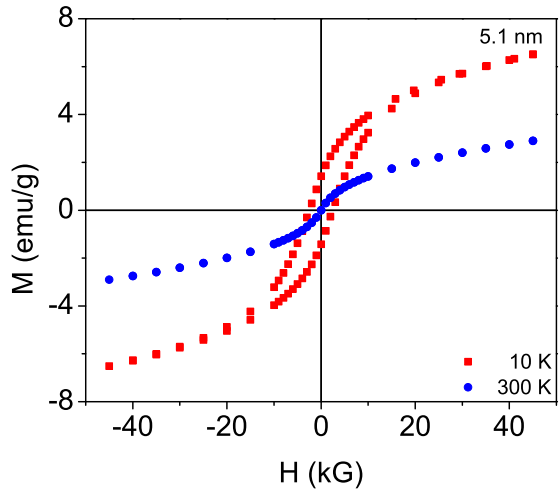}
\caption{(color online) M-H curves at 10~K and 300~K for the 5.1~nm particles.  Hysteresis is seen in the 10~K data while there is no hysteresis at 300~K.}
\label{fig:dc-mag-hysteresis}
\end{center}
\end{figure}

In Figure \ref{fig:dc-mag-hysteresis} we show the M-H curves obtained from \emph{dc} magnetization measurements at 10~K and 300~K for the 5.1~nm particles. Hysteresis loop is seen in the 10~K data. The hysteresis could be an indication of the weak ferromagnetism alluded to by N\'{e}el in his musings on antiferromagnetic particles \cite{Low Temp. Phys.}; or it could be because the magnetization is time dependent and is relaxing too slowly for the experimental time scale of a few hours. We shall soon see that our data support the latter proposition. At 300~K there is no coercive force and no hysteresis. We carried out a few more $M$ vs. $H$ measurements at different temperatures up to 350~K  and found that the magnetization is not a function of $H/T$ as one would expect for superparamagnetic systems.  Similar behavior has been reported by others also on NiO nanoparticles \cite {Makhlouf 1997}.

\begin{figure} [hbt]
\begin{center}
\includegraphics[angle=0,width=.45\columnwidth]{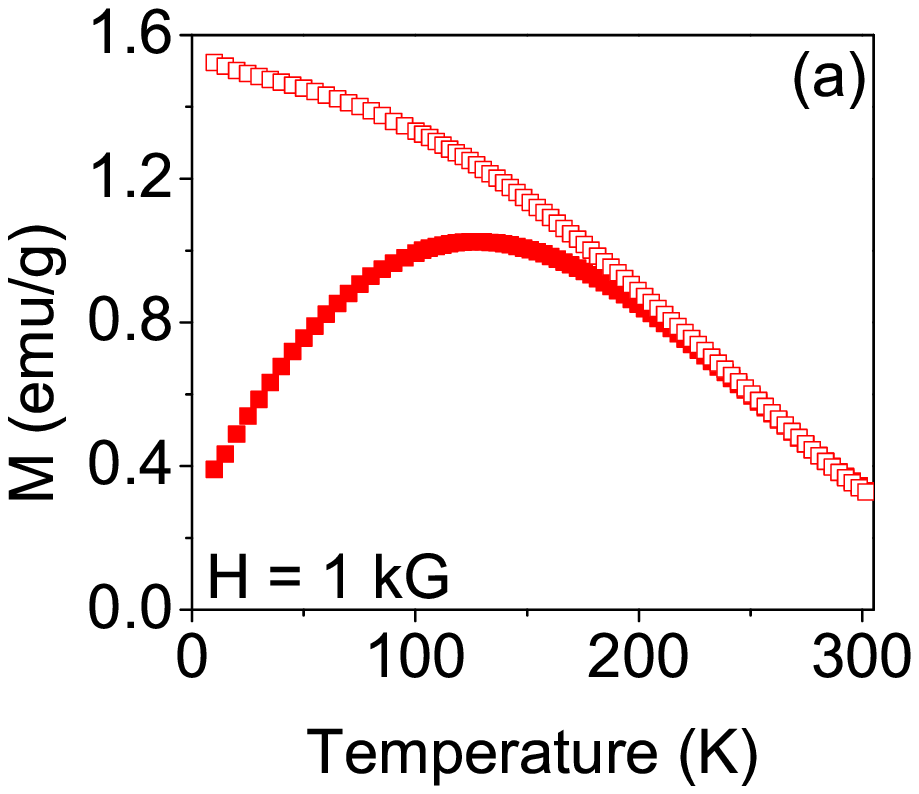}
\includegraphics[angle=0,width=.45\columnwidth]{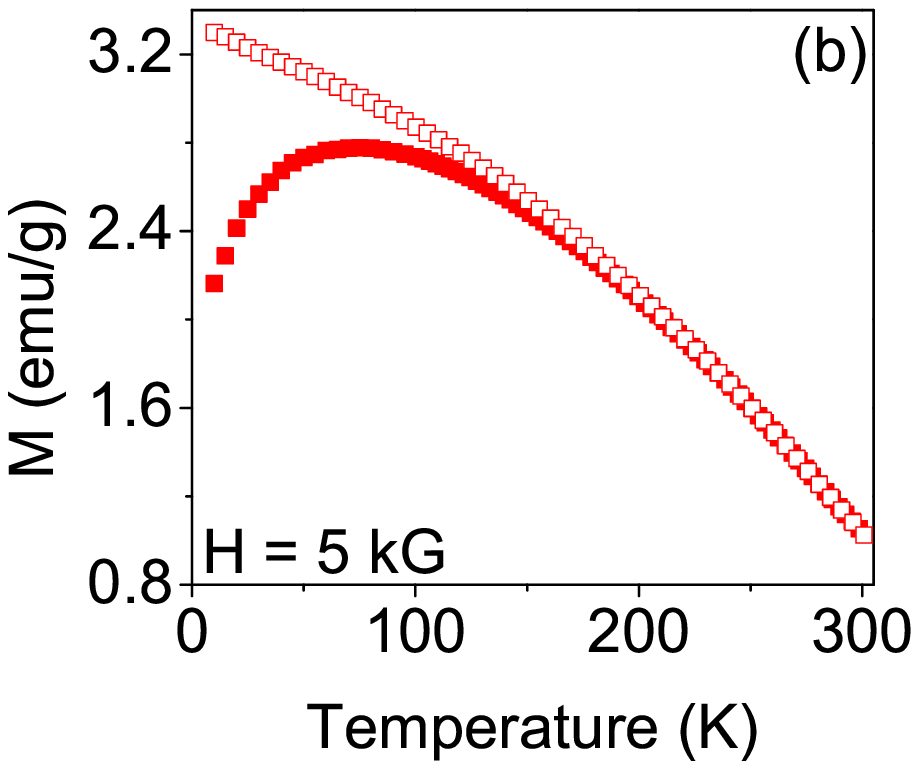}
\end{center}
\begin{center}
\includegraphics[angle=0,width=.45\columnwidth]{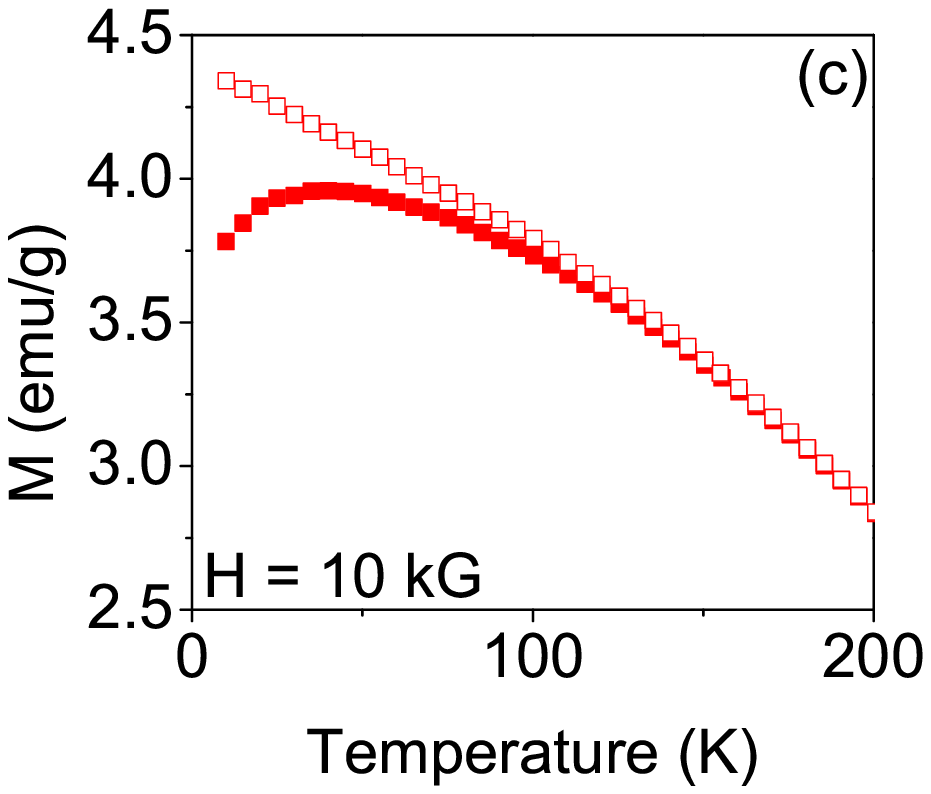}
\includegraphics[angle=0,width=.45\columnwidth]{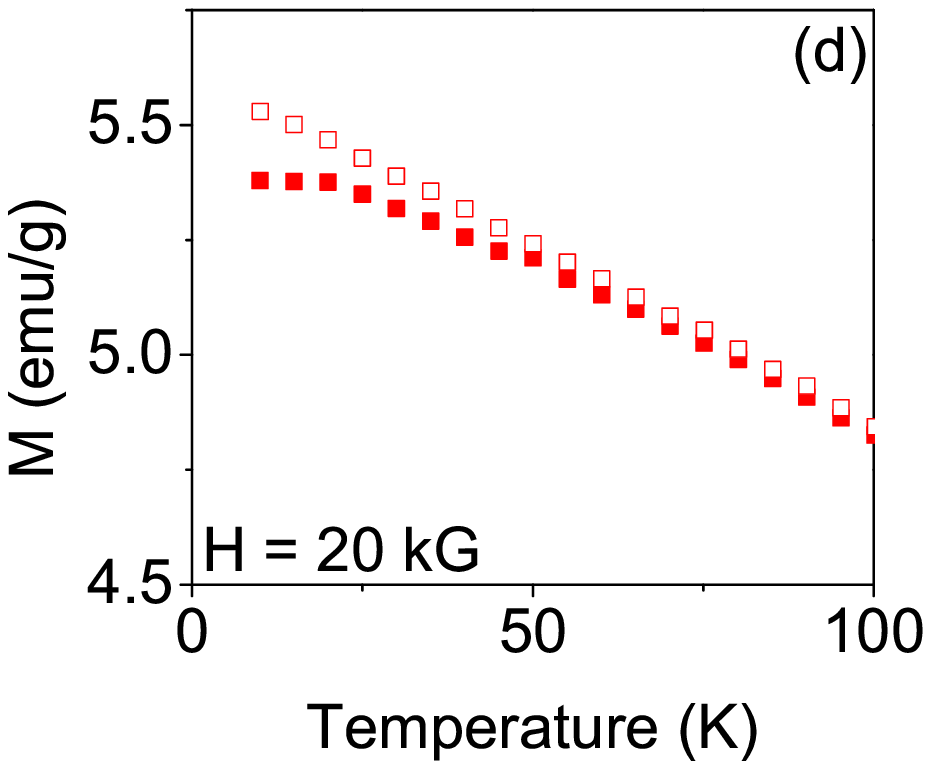}
\end{center}
\caption{(color online) ZFC (solid symbol)\ and FC (open symbol)\ magnetization as a function of temperature in different high fields. Please note that the scales are different for each graph.}
\label{fig:dc-mag-FC-ZFC}
\end{figure}

In Figure \ref{fig:dc-mag-FC-ZFC} we show the ZFC and FC magnetization curves for the 5.1~nm sample at various applied fields. As the measuring field increases we find that both the bifurcation temperature of ZFC and FC curves and the peak temperature of ZFC curve shift to low temperature. In fact, at 20 kG the ZFC curve does not peak even on going down to 10~K.  This behavior is along expected lines since we would expect in a sufficiently high external magnetic field the spin glass freezing will either take place at a lower temperature or not at all as a sufficiently strong external  magnetic field can break the spin glass phase.

\subsubsection{Time Dependence}
\label{subsec:Time-depend-TRM}

Thermoremanent magnetization (TRM) is the time dependent remanent magnetization obtained when a sample is cooled from a temperature well above $T_{f}$ to a temperature below $T_{f}$ in  an applied field $H$ and subsequently the field is removed. In this work, to measure the TRM, we cool the sample from 300~K to the temperature of interest in a 100~G magnetic field.

A spin glass system is not in thermal equilibrium and slowly relaxes to lower energy states. Many models have been proposed to describe the time dependent relaxation of magnetization in spin glasses. A single energy barrier should give an exponentially relaxing TRM, with time constant $\tau$, ie., $M(t) = M_{0} \exp{(-(t/\tau))}$. In a real system there would be a range of energy barriers and hence there would be a distribution of relaxation times. One of the popular forms incorporating a distribution of relaxation times is the stretched exponential function: \cite{Hoogerbeets, Chamberlin}

\begin{equation}
\label{eqn:stretched-exp}
M(t) = M_{0} \exp{(-(t/\tau)^{n})} 
\end{equation}
where $\tau$ is a characteristic relaxation time and $M_{0}$ and $n$ are fit parameters.

By assuming that the energy barriers are uniformly distributed from zero to some maximum energy, it has been shown that the magnetization decays logarithmically  \cite{Guy} as

\begin{equation}
\label{eqn:log-decay}
M(t) = M_{0} - S \log(t)
\end{equation}
where $M_{0}$ and $S$ are fit parameters. It should be noted that $M_0$ is the magnetization at $t=1$ unit and hence it depends on the unit of time used, while $S$ itself does not have any such dependence.

A power law form has also been used to describe the decay of magnetization: \cite{G.Sinha, Patel}

\begin{equation}
\label{eqn:power-law}
M(t) = M_{0} t^{-n}
\end{equation}
where the exponent $n$ should increase with increasing temperature. In this case also, as in the previous equation, $M_0$ depends on the unit of time.

A statistical measure of the goodness of a fit is $\chi^{2}$, defined as \cite{Press}
\begin{equation}
\chi^{2} = \sum_{i=1}^{N} \frac{{[M_i (measured) - M_i (fit)]}^{2}}{\sigma_{i}^{2}}
\end{equation}
where N is the total number of data points and $\sigma_{i}$ is the standard deviation of the $i^{th}$ data point. For a good fit $\chi^{2}$ should have a value close to N. We shall also be reporting the coefficient of determination $R^2$ for our fits; the closer $R^2$ is to unity the better the fit. 

We fitted our data to the Equations (\ref{eqn:stretched-exp}), (\ref{eqn:log-decay}) and (\ref{eqn:power-law}). To compare the fits for the different forms we present the values of $\chi^{2}$ and $R^2$ for a representative data set, the 100~K TRM data for 5.1~nm particles, in Table \ref{table-comparison-of-forms}. We note that the stretched exponential gives the best fit with the lowest $\chi^{2}$ and largest $R^2$. This is followed by the logarithmic form with the power law form bringing up the rear. To check whether the fit parameters truly characterize the data we decided to run the fits on subsets of the original data covering different time spans such as (150 s, 1000 s), (150 s, 2000 s), (150 s, 3000 s) and so on. The fit parameters are found to be strongly dependent on the time span of the data for the stretched exponential case but are found to be almost independent of the same for logarithmic and power law cases. Consequently the stretched exponential fit has to be rejected on the grounds that the fit parameters in that case are simply artifacts of the subset of the data used, even though it has the best $\chi^{2}$  and $R^2$ values.

\begin{table}
\caption{Values of $\chi^{2}$ and the coefficient of determination $R^{2}$ obtained by fitting 100~K TRM data to various expressions for 5.1~nm sample. Total number of data points, $N$, is 749.}
\label{table-comparison-of-forms}
\vspace{0.2in}
\begin{tabular}{|c|c|c|}
\hline 
Fit Expression&$\chi^{2}$&$R^{2}$\\
\hline 
$M_{0} \exp{(-(t/\tau)^{n})}$&810&0.9998\\
$M_{0} - S \log(t)$&1008&0.9997\\
$M_{0} t^{-n}$&2547&0.9992\\
\hline
\end{tabular}
\end{table}

\begin{table}
\caption{Values of fit parameters $M_{0}$ and $S$ to Equation (\ref{eqn:log-decay}) and the values of $\chi^{2}$ and  $R^{2}$ for 5.1~nm sample at different temperatures. Total number of data points, $N$, is 749.}
\label{table-fit-parameters}
\vspace{0.2in}
\begin{tabular}{|c|c|c|c|c|}
\hline 
T&$M_{0}$&$S$&$\chi^{2}$&$R^{2}$\\
(K)&($10^{-2}$emu/g)&($10^{-3}$emu/g)& &\\
\hline 
25&20.7&2.94&1219&0.9898\\
50&17.7&4.46&993&0.9979\\
75&14.7&6.20&1500&0.9991\\
100&11.9&7.06&1008&0.9997\\
125&8.34&6.29&2996&0.9995\\
150&5.54&5.02&1259&0.9998\\
175&3.32&3.56&10170&0.9993\\
200&1.73&2.32&32400&0.9991\\
\hline
\end{tabular}
\end{table}

\begin{figure}[htb]
\begin{center}
\includegraphics[angle=0,width=1.\columnwidth]{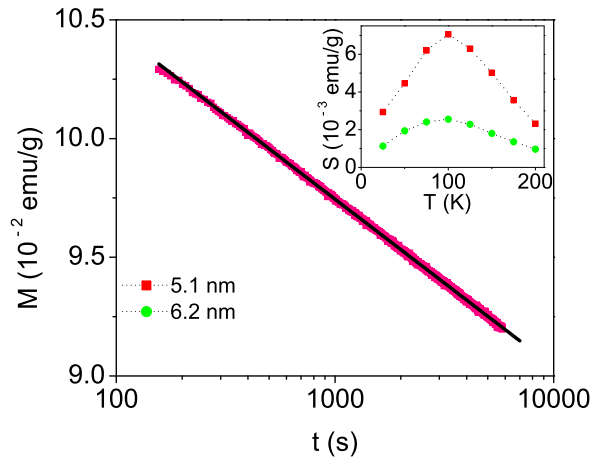}
\caption{(color online) Time decay of the thermoremanent magnetization of 5.1~nm particles at 100~K.
The solid line represent the fit to the Equation (\ref{eqn:log-decay}). The inset shows $S$ as a function of temperature}
\label{fig:time-dependence-log-fit}
\end{center}
\end{figure}

Now we are left with the logarithmic and power law forms of which the logarithmic form has the better $\chi^{2}$ and $R^2$. We found that Equation (\ref{eqn:log-decay}) gives the best fit not only at 100~K, but at all the other temperatures as well. We conclude that the best fit to our TRM data is given by this equation. This equation  has been popularly used to describe the time dependence of TRM of various spin glass systems including bulk materials \cite{Guy, Chun}, nanoparticles \cite{Luo} and thin films \cite{Dhar}. In Table \ref{table-fit-parameters} we present the results of fitting the TRM data to the logarithmic form. We see that the $\chi^{2}$ is of the order of $N$  and the $R^{2}>0.999$ in most of the cases which reflect the high quality of the fits. It can be seen from the table that as the temperature increases the parameter $M_0$ decreases monotonically as is to be expected. In Figure \ref{fig:time-dependence-log-fit} we show the thermoremanent magnetization of the 5.1~nm particles at 100~K along with the logarithmic fit. The solid line through the data points is the fit and we note that it is excellent. In the inset we plot the fit parameter $S$, which characterizes the relaxation rate, as a function of temperature for the 5.1~nm and 6.2~nm particles. We see that $S$ peaks around 100~K for both samples. Peaks in $S$ vs. $T$ curves have been seen in other spin glass systems also\cite{Guy, Luo}.

Effect of wait time dependence on TRM for spin glasses has been studied by several groups \cite{Chamberlin PRB, Hoogerbeets, Nordblad}. It is generally observed that the rate of magnetization decay decreases with increasing wait time. We also studied the wait time dependence of decay of TRM at 100~K for the 5.1~nm system. For this the sample is cooled from 300~K to 100~K in a field of 100~G and then we wait for a time $t_w$ before the field is switched off and the data acquisition is started. In Table \ref{table-wait-time} we show the fit parameters to Equation (\ref{eqn:log-decay}). It is clear that there is a small decrease in the fit parameter $S$ for increasing wait time up to 1 hour which is consistent with other's observations \cite{Chamberlin PRB, Nordblad, Lundgren}. 

\begin{table}
\caption{Fit parameters $M_{0}$ and $S$  for 5.1~nm sample for different wait times $t_w$ at 100~K.}
\label{table-wait-time}
\vspace{0.2in}
\begin{tabular}{|c|c|c|c|c|}
\hline 
$t_w$&$M_{0}$&$S$&$R^{2}$\\
(minutes)&($10^{-2}$emu/g)&($10^{-3}$emu/g)&\\
\hline 
0 &11.9&7.06&0.9997\\
15&11.8&6.99&0.9994\\
30&11.7&6.92&0.9994\\
60&11.7&6.91&0.9991\\
\hline
\end{tabular}
\end{table}

\subsection{Further Discussion}
\label{Further Discussion}
In Sections \ref{subsec:ac-suscept-particle-size} and \ref{subsec:dc-mag-particle-size} we saw that the peak temperature of the low field susceptibility as a function of temperature decreases with increasing particle size. We take this to mean that the behavior of the system is not superparamagnetic and the peaks the in low field susceptibility vs. temperature curves are not because of superparamagnetic blocking of particle magnetic moments. From Sections \ref{subsec:T&Fdepend-ac-chi}, \ref{subsec:Field-depend-ac-chi} and \ref{subsec:Time-depend-TRM} it is clear that the system is showing spin glass behavior. Now the question is, what is the mechanism behind the spin glass freezing? 

There are reports in the literature where dipolar interaction between particles has been proposed as the reason for the freezing of particle magnetic moments \cite{Morup 1995, Morup 1983, Luo}. Now let us examine whether there is such a possibility in the case of NiO. We estimate that for the 5.1~nm particles there would be an average uncompensated moment of the order of $100 \mu_{B}$ \cite{Note1}. The maximum dipolar interaction energy between two such particles touching each other would be $\sim 10^{-17}$ erg which corresponds to about 0.1~K on temperature scale. This means that if dipolar interaction were causing the freezing it would occur at about 0.1~K which is much lower than the observed freezing temperature of about 167~K (from $ac$ susceptibility). Thus we rule out the possibility that dipolar interaction among particles is causing the peaks in low field susceptibility as a function of temperature.


\begin{figure}[tb]
\begin{center}
\includegraphics[angle=0,width=1.\columnwidth]{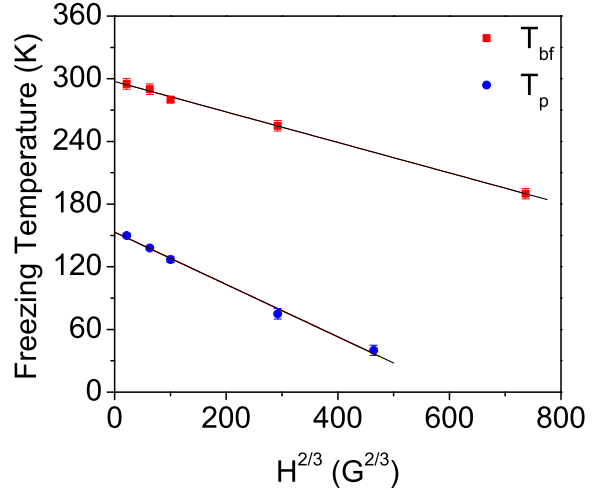}
\caption{(color online) The point of bifurcation ($T_{bf}$) of ZFC and FC magnetization curves is considered to be the starting point of the freezing process. The peak temperature ($T_p$) in the ZFC magnetization curve is generally considered  as the freezing temperature. This figure shows plots of $T_{bf}$ and $T_p$ vs. $H^{2/3}$. The solid lines show linear fits to the data with $R^2$ of 0.9984 and 0.9954 respectively. The upper and lower curves have zero field intercepts of  297~K and 153~K respectively.}
\label{fig:high-field-AT-line}
\end{center}
\end{figure}

In canonical spin glasses the spin glass phase is very sensitive to external fields and the application of a field of a few hundred gauss breaks it and makes it disappear. But here it is not the case and the spin glass phase survives up to the highest field used to measure ZFC and FC magnetization  (up to 20 kG in our study; see Figure \ref{fig:dc-mag-FC-ZFC}). The existence of spin glass phase up to such high fields is also evident from the AT lines shown in Figure \ref{fig:high-field-AT-line}. Clearly the peak temperature of the susceptibility curve decreases with increasing applied field following the AT line. We also find that the rate of change with field of the peak temperature of the susceptibility curve decreases with increasing particle size (not shown here). To paraphrase, for smaller particles the $T_p$ changes more with field than for the larger particles. It is this which gives rise to and resolves the conflicting observations noted at the end of Section \ref{subsec:dc-mag-particle-size}.

Kodama et al. \cite{Kodama 1996} and Mart\'{i}nez et al. \cite{Martinez} have argued that freezing of surface spins can lead to spin glass phase which survives up to such high fields. Kodama et al. \cite{Kodama 1996} have proposed a model for ferrimagnetic nickel ferrite nanoparticles where the spins at the surface of a particle are disordered, leading to frustration and spin glass like freezing. A similar model has also been proposed for  antiferromagnetic ferrihydrite nanoparticles \cite{Seehra 2000}. We feel that such a model may be applicable in the case of NiO nanoparticles as well. We will first have to see whether we can propose any mechanism for surface spin disorder in our case. The exchange interaction between two neighboring Ni$^{+2}$ ions is mediated by an oxygen ion (superexchange), and if an oxygen ion is missing from the surface, the exchange bond would be broken and the interaction energy would be reduced. Also, the average coordination number  for Ni$^{+2}$ ions at the surface will be less than that in the bulk and this can result in a distribution of exchange energies for the surface spins. Moreover, the superexchange is sensitive to bond angles and bond lengths, which are likely to be different at the surface compared to that in the bulk. The reasons mentioned above may be sufficient to give rise to surface spin disorder and frustration leading to a spin glass phase. The fraction of atoms lying on the surface of the particles increases with decreasing particle size. This may lead to increased surface spin disorder as the particle size decreases and could possibly account for the increasing freezing temperature with decreasing particle size.         

\section{Conclusion}
In this paper we reported a detailed study on magnetic properties of sol-gel prepared NiO nanoparticles. The behavior of the system is not superparamagnetic as was expected. In fact it shows spin glass behavior which we attribute to surface spin disorder.

\acknowledgments

We thank Profs. A. K. Majumdar, Avinash Singh and R. C. Budhani for helpful discussions.


\begin{thebibliography}{6}

\bibitem{Chantrell} R. W. Chantrell and K. O'Grady, in \emph{Applied Magnetism}, edited by R. Gerber et al. (Kluwer Academic Publishers, The Netherlands, 1994), p.113.

\bibitem{Neel} L. N\'{e}el, Ann. Geophys \textbf{5}, 99 (1949).

\bibitem{Brown} W. F. Brown, Jr., Phys. Rev. \textbf{130}, 1677 (1963).

\bibitem{Andersson} J. O. Anderson, C. Djurberg, T. Jonsson, P. Svedlindh, and P. Nordblad, Phys. Rev. B \textbf{56}, 13983 (1997); C. Djurberg, P. Svedlindh, P. Nordblad, M. F. Hasen, F. B{\o}dker, and S. M{\o}rup, Phys. Rev. Lett. \textbf{79}, 5154 (1997); J. L. Dormann, D. Fiorani, R. Cherkaoui, E. Tronc, F. Lucari, F. D'Orazio, L. Spinu, M. Nogu\`{e}s, H. Kachkachi, and J. P. Jolivet, J. Magn. Magn. Mater. \textbf{203}, 23 (1999); J. L. Dormann, D. Fiorani, R. Cherkaoui, L. Spinu, F. Lucari, F. D'Orazio, M. Nogu\`{e}s, E. Tronc, J. P. Jolivet, and A. Garcia, Nanostructured Materials \textbf{12}, 757 (1999). 

\bibitem{Low Temp. Phys.} L. N\'{e}el, in \emph{Low Temperature Physics}, edited by C. Dewitt et al. (Gordan and Beach, New York, 1962), p.413.

\bibitem{Seehra 2000} M. S. Seehra, V. S. Babu, A. Manivannan, and J. W. Lynn, Phys. Rev. B \textbf{61}, 3513, (2000).

\bibitem{Harris} J. G. E. Harris, J. E. Grimaldi, D. D. Awschalom, A. Chiolero, and D. Loss, Phys. Rev. B \textbf{60}, 3453 (1999);   S. M{\o}rup and C. Frandsen, Phys. Rev. Lett. \textbf{92}, 217201 (2004).

\bibitem{Kilcoyne} S. H. Kilcoyne and R. Cywinski, J. Magn. Magn. Mater. \textbf{140-144}, 1466 (1995). 

\bibitem{Makhlouf} S. A. Makhlouf, F. T. Parker, and A. E. Berkowitz, Phys. Rev. B \textbf{55}, R14717 (1997).

\bibitem{Smart} J. S. Smart and S. Greenwald, Phys. Rev. \textbf{82}, 113 (1951).

\bibitem{Richardson and Milligan} J. T. Richardson and  W. O. Milligan, Phys. Rev. \textbf{102}, 1289 (1956).

\bibitem{Makhlouf 1997} S. A. Makhlouf, F. T. Parker, F. E. Spada, and A. E. Berkowitz, J. Appl. Phys. \textbf{81}, 5561 (1997).

\bibitem{Richardson 1991} J. T. Richardson, D. I. Yiagas, B. Turk, K. Foster, and M. V. Twigg, J. Appl. Phys. \textbf{70}, 6977 (1991). 

\bibitem{Scherrer} B. D. Cullity, \emph{Elements of X-Ray Diffraction} (Addison-Wesley Publishing Company, Inc., 1956),  p.99 \& 262.

\bibitem{arxiv} S. D. Tiwari and K. P. Rajeev, www.arXiv.org/cond-mat/0408427 (2004).

\bibitem{Jonsson} T. Jonsson, P. Nordblad, and P. Svedlindh, Phys. Rev. B \textbf{57},
 497 (1998).

\bibitem{Mydosh} J. A. Mydosh, \emph{Spin Glass} (Taylor \& Francis, 1993), p.65-67.

\bibitem{Bean} I. S. Jacobs and C. P. Bean, in \emph{Magnetism},Vol.III edited by G. T. Rado and H. Suhl (Academic Press Inc., New York, 1963), p.271.


\bibitem{Bitoh} T. Bitoh, K. Ohba, M. Takamatsu, T. Shirane, and S. Chikazawa, J. Phys. Soc. Jpn. \textbf{64}, 1305 (1995).

\bibitem{Labarta} A. Labarta, O. Iglesias, Ll. Balcells, and F. Badia, Phys. Rev. B \textbf{48}, 10240 (1993).

\bibitem{Morup 1995}S. M{\o}rup, F. B{\o}dker, P. V. Hendriksen, and S. Linderoth, Phys. Rev. B \textbf{52}, 287 (1995).

\bibitem{AT} J. R. L. de Almeida and D. J. Thouless, J. Phys. A \textbf{11}, 983 (1978).

\bibitem{Chamberlin 1982} See for instance: R. V. Chamberlin, M. Hardiman, L. A. Turkevich, and R. Orbach, Phys. Rev. B \textbf{25}, 6720 (1982); P. Monod and H. Bouchiat, J. Phys. (Paris) Lett. \textbf{43}, 145 (1982).

\bibitem{Martinez} B. Mart\'{i}nez, X. Obradors, Ll. Balcells, A. Rouanet, and C. Monty, Phys. Rev. Lett. \textbf{80}, 181 (1998). 

\bibitem{Dhar} S. Dhar, O. Brandt, A. Trampert, K. J. Friedland, Y. J. Sun, and K. H. Ploog, Phys. Rev. B \textbf{67}, 165205 (2003).


\bibitem{Luo} W. Luo, S. R. Nagel, T. F. Rosenbaum and R. E. Rosensweig, Phys. Rev. Lett. 67, 2721 (1991).

\bibitem{Seo} W. S. Seo, H. H. Jo, K. Lee, B. Kim, S. J. Oh, and J. T. Park, Angew. Chem. Int. Ed. \textbf{43}, 1115 (2004).


\bibitem{Hoogerbeets} R. Hoogerbeets, W. Luo, and R. Orbach, Phys. Rev. Lett. \textbf{55},111 (1985).

\bibitem{Chamberlin} R. V. Chamberlin, G. Mozurkewich, and R. Orbach, Phys. Rev. Lett. \textbf{52}, 867 (1984).



\bibitem{Guy} C. N. Guy, J. Phys. F: Metal Phys. \textbf{8}, 1309 (1978).

\bibitem{G.Sinha} G. Sinha, R. Chatterjee, M. Uehara and A. K. Majumdar, J. Magn.Magn.Mater. \textbf{164}, 345 (1996). 


\bibitem{Patel} R. S. Patel, D. Kumar, and A. K. Majumdar, Phys. Rev. B \textbf{66}, 54408 (2002).


\bibitem{Press} W. H. Press et al., \emph{Numerical Recipes in C} (Cambridge University Press, 1992), p. 659.

\bibitem{Chun} S. H. Chun, Y. Lyanda-Geller, M. B. Salamon, R. Suryanarayanan, G. Dhalenne, and A. Revcolevschi, J. Appl. Phys. \textbf{90}, 6307 (2001). 




\bibitem{Nordblad} P. Nordblad, P. Svedlindh, L. Lundgren, and L. Sandlund, Phys. Rev. B \textbf{33}, 645 (1996).


\bibitem{Chamberlin PRB} R. V. Chamberlin, Phys. Rev. B \textbf{30}, 5393 (1984).

\bibitem{Lundgren} L. Lundgren, P. Svedlindh, P. Nordblad, O. Beckman, Phys. Rev. Lett. \textbf{51}, 911 (1983).


\bibitem{Morup 1983} S. M{\o}rup, M. B. Madsen, J. Franck, J. Villadsen, and C. 
J. W. Koch, J. Magn. Magn. Mater. \textbf{40}, 163 (1983); I. Tamura and M. Hayashi, J. 
Magn. Magn. Mater. \textbf{72}, 285 (1988); D. H. Reich, B. Ellman, J. Yang, T. F. Rosenbaum,
G. Aeppli, and D. P. Belanger, Phys. Rev. B \textbf{42}, 4631 (1990).

\bibitem{Note1} It has been reported that 15~nm particles of NiO prepared by the same method as ours has a magnetic moment of $700 \mu_{B}$ [See R. H. Kodama, S. A. Makhlouf, and A. E. Berkowitz, Phys. Rev. Lett. \textbf{79}, 1393 (1997)]. In the case of antiferromagnetic particles the moment $\propto (particle size)^{n}$ where n should be a number between 1 and 2 (see ref. \cite{Low Temp. Phys.} and \cite{Richardson 1991}). Taking a value of $n \sim 1.5$, for the 5.1~nm sized particles, we estimate that, the moment should be $\sim 100 \mu_{B}$.

\bibitem{Kodama 1996} R. H. Kodama, A. E. Berkowitz, E. J. McNiff, Jr., and S. Foner, Phys. Rev. Lett. \textbf{77}, 394 (1996).




\end{thebibliography}
\end{document}